\documentclass[preprint2]{aastex}
\usepackage{emulateapj5}
\usepackage{epsfig}
\usepackage{onecolfloat5}
\newenvironment{figurehere}{\begin{figure}[htbp]\begin{center}}{\end{center}\end{figure}}

\def\apj{ApJ}

\def\mnras{MNRAS}

\def\nat{Nature}

\newcommand{\vect}[1]{\ensuremath{\mbox{\boldmath $#1$}}}

\def \retilde {{\tilde r_{\rm E}}}
\def \thetae {\theta_{\rm E}}
\def \pie {\vect{\pi}_{\rm E}}
\def \dol {D_{\rm ol}}
\def \dls {D_{\rm ls}}
\def \dos {D_{\rm os}}
\def \earth {\oplus}
\def \kms {\rm km \, s^{-1}}
\def \teff {T_{\rm eff}}
\def \u {\vect{u}}
\def \tint {t_{\rm int}}

\lefthead{Graff \& Gould}
\righthead{Microlens Parallaxes of Binary Lenses}
\submitted{Submitted to ApJ March 19, 2002}

\begin{document}
\singlespace
\twocolumn[

\title{Microlens Parallaxes of Binary Lenses\\
Measured from a Satellite}

\author{David S. Graff and Andrew Gould}
\affil{The Ohio State University Department of Astronomy, Columbus, OH
  43210\\
graff,gould@astronomy.ohio-state.edu}
\begin{abstract}
Caustic-crossing binary lenses make up about 5\% of all detected
microlenses.  The relative proper motion of a caustic-crossing binary
lens can be measured with observations from a single terrestrial
telescope.  Thus, uniquely, binary lenses can be completely solved
with only the addition of a measurement of the microlensing parallax.
This solution will yield the mass, distance, and transverse velocity
of the lens relative to the source.  To date, only one of the $\sim
1000$ observed microlensing events has been so solved.

We examine the ability of a parallax satellite combined with
ground-based observations to solve these events.  To measure both
components of the vector parallax, the lens must be observed near two
different caustics.  Thus, the final accuracy is determined mostly by
whether one can intensively monitor part of the first caustic
crossing, by the magnification pattern, and by the path of the source
with respect to the lens geometry.  We find that vector parallaxes can be
measured far more easily for binary lenses than single lenses,
requiring 1-3 orders of magnitude fewer photons.  They may thus yield
a large number of completely solved lenses relatively cheaply.
\end{abstract}

]
\section{Introduction}

In any microlensing event, there are four lens parameters of interest,
its mass, $M$, the two components of its proper motion relative to the
source, $\vect{\mu}_{\rm rel}$, and its parallax relative to the
source, $\pi_{\rm rel}$.  To measure these four parameters, one must
measure four independent quantities.  However, in standard
microlensing, one measures only a single quantity of physical
interest, the Einstein radius crossing time, $t_{\rm E}$.  Thus, for
the vast majority of microlensing events, one knows only a single
degenerate combination of the four parameters, and can thus say very
little about the lens.  The individual parameters must be
statistically inferred from a Galactic model.  To date, the
interpretation of microlensing results along every line of sight is
subject to fierce controversy, as, in general, no standard Galactic
model can explain any of the results.  Fully solving microlensing
events would allow us to measure the mass function of the Galactic bulge, study
the spatial and kinematic structure of the Galaxy, and possibly
determine the nature and location of at least a major component of
Galactic dark matter.

The three other observable quantities that must be measured to solve an
event are the angular Einstein radius, $\thetae$, the Einstein radius
projected on the observer plane, $\retilde$ and the direction of the
lens motion, $\alpha$.  All the lens parameters can be simply expressed
in terms of these quantities \citep{g00a}.  For example, the lens mass
is given by
\begin{equation}
M = {c^2\over 4 G} \thetae \retilde \, .
\end{equation}

Of the roughly 1000 microlensing events observed to date, there are
measurements of ${\retilde}$ for only about a dozen
\citep{machoretilde,b01,m99,s01,OGLEII,moa01,ogle02,an02}, and measurements
of $\thetae$ for a similar number
\citep{macho97,macho00a,machonature,a99a,mb41,ob23,smc,an02}.  Typically,
the events with measured $\retilde$ are the easiest few to measure,
with longer than average time scales, and thus they do not
characterize the lens population.  Moreover, for only one of these
\citep{an02} was it possible to measure both ${\retilde}$ and
$\thetae$, and so to completely solve the event.  To routinely solve
typical microlensing events, and thereby measure $M$ and other useful
lens parameters, one must be able to routinely measure both $\retilde$
and $\thetae$.  The projected Einstein radius ${\retilde}$ can be
measured by comparing photometry of the event from the Earth and a
satellite in solar orbit \citep{r66,g94,g95a}.  The angular Einstein
radius $\thetae$ can be measured by tracking the excursion of the
centroid of the lens images relative to the position of the source
\citep{hnp95,w95,my95}.  For typical events the scale of this
excursion is only $\sim 100\,\mu$as, implying that only with
space-based astrometric interferometers are accurate measurements
feasible \citep{p98,bsvb}.

All studies of astrometric mass measurements to date have considered
only microlensing by single lenses, not binaries.  At first this
appears to be a very reasonable simplification because, while the
majority of stars reside in binaries \citep{dm91}, when the binary
angular separation is much larger or smaller than $\thetae$, binary
microlensing can hardly be distinguished from single-lens
microlensing.  Indeed, \citet{macho00b} and \citet{u00} found that
only $\sim 4-6.5\%$ respectively of observed events are caustic
crossing binaries.  A somewhat higher number are detected as
binaries, but do not cross caustics, and a much larger number are
undoubtedly binaries, but do not show significant deviation from a
single lens lightcurve \citep{distephano}. To a first approximation, it
therefore seems that not much is lost by ignoring this $4-6.5\%$.

In fact, this oversight is quite important.  There are two key ways in
which binary lenses are far easier to solve than single lenses: they require
many fewer photons to determine ${\retilde}$, and $\thetae$ can be
determined without having to resort to astrometry.

First, we will show that for fixed source brightness, ${\retilde}$
can be determined for a binary microlens with only about 1\% of the
observing time required for single lenses.  This means that $I=17.5$
binary events can be measured in only one tenth the time needed for
$I=15$ single-lens events.  And while $I=15$ single-lens events are 15
times more frequent than $I=15$ binary events, they have roughly the
same frequency as $I=17.5$ binary events.  Hence, binary-lens events
allow one to greatly increase the total number of measurements at a very
modest observing cost.
	
Second, $\thetae$ can be independently measured from the ground in
caustic-crossing binary lenses.  This has been almost the only
technique by which $\thetae$ has been measured to date
\footnote{The
exceptions are MACHO 95-BLG-30 for which the source crossed the point
caustic at the center of a single lens instead of the fold-caustic of
a binary lens \citep{macho97}, and MACHO LMC-5 for which $\thetae$ was
measured with LMC astrometry \citep{machonature}.}.  
The finite disk
of the source star takes time $2 \Delta t$ to cross the caustic, which
can be measured directly from the lightcurve.  This time is related
to $\thetae$ by,
\begin{equation}
\frac{\Delta t}{t_{\rm E}} \cos{\phi} =  \frac{\theta_*}{\thetae} \, ,
\end{equation}
where $\theta_*$ is the angular size of the source star, and $\phi$ is
the angle of the source trajectory with respect to the normal to the
caustic.  The source star size, $\theta_*$ can be determined from its
(dereddened) flux and effective temperature by
\begin{equation}
\label{teff}
{\cal F} =  \theta_*^2 \sigma \teff^4 \, ,
\end{equation} 
which relation has been best calibrated by \citet{vanbelle} using
$(V-K)$ as a probe of surface temperature.  Thus, if there is no high
precision astrometry to determine $\thetae$, binary lenses will be essentially the
only lenses that can be completely solved from parallax measurements.

\section{Theory}

The Einstein angle of a binary lens is given by,
\begin{equation}
\thetae^2 =  \frac{4G(M_1+M_2)}{Dc^2},
\qquad D\equiv \frac{\dol \dos}{\dls}\, .
\end{equation}
Here, $M_{1,2}$ are the lens masses, and $\dls$, $\dos$, and $\dol$
are the distances between the observer, lens, and source.  The
projected Einstein radius,
\begin{equation}
\retilde = D \thetae \,
\end{equation}
defines the scale of the magnification pattern projected onto the
observer plane.  The position of the observer in this plane in units
of $\retilde$ is denoted $\u$.

Note that $\u\thetae$ is the angular displacement from the lens to the
source as seen from an observer located at $\u\retilde$ in the
observer plane.  Thus, we caution the reader not to get confused that
we will use $\u$ to refer both to the position of the source and the
position of the observer.  When discussing observations from a single
telescope, which has characterized the situation in the vast majority
of the microlensing literature, it is usually more convenient to think
of a fixed observatory and a source moving behind the lens at location
$\u\thetae$.  When discussing simultaneous observations from several
telescopes distributed about the solar system, it is more convenient
to think of a group of telescopes moving through a fixed magnification
pattern with individual telescopes located at $\u\retilde$.  The two
frames are perfectly consistent and interchangeable.

The photometric magnification is a function of $\u$, $A(\u)$.
The angular separation of the two elements of the lens is defined to
be $\vect{d} \thetae$ with $\vect{d}$ pointing from the primary to the
secondary.  We will fix the origin of the $\u$ plane at the midpoint
between the two stars in the binary.  The two stars of the lens are
thus located at $\pm \vect{d}/2$.  It is conventional to align the
coordinates of the $\u$ plane along $\vect{d}$.

The Sun moves through the observer plane with rectilinear motion:
\begin{equation}
\u_\odot(t) = \u_{0,\odot} + \vect{\mu}_\odot (t - t_0) \, .
\end{equation}
while the other objects in the solar system are displaced from this
position by their actual positions in the solar system projected along
the line of sight, and brought into scale by dividing by $\retilde$.
For example, the position of the Earth is
\begin{equation}
\u_\earth = \u_\odot - \frac{\vect{\hat{n}} \times \vect{\hat{n}} \times
\vect{a}_{\earth \odot}(t)} {\retilde}
\end{equation}
where $\vect{\hat{n}}$ is the unit vector in the direction of the source and
$\vect{a}_{\earth \odot}(t)$ is the displacement between the Earth and
Sun.  In this paper, we only consider observations over a short
period of time, about 1 month.  We will thus ignore the parallax effect of the
Earth's motion around the Sun, and model the Earth's motion as rectilinear,
\begin{equation}
\u_\earth(t) = \u_{0,\earth} + \vect{\mu}_\earth (t - t_0) \, .
\end{equation}

We shall assume that there are two telescopes
monitoring the event, say, one on the Earth and one on a satellite.
The question before us is: how closely will one be able to determine
the positions of the telescopes in the observer plane based on the
magnifications recorded by the two telescopes?  These two positions,
that of Earth and the satellite, are related by
\begin{equation}
\label{retildeeq}
\u_\earth - \u_s  \equiv \delta \u = \frac{\vect{\hat{n}} \times \vect{\hat{n}} \times
\vect{a}_{\earth s}(t)} {\retilde} \, .
\end{equation}

We see from equation (\ref{retildeeq}) that $\retilde$ is inversely
proportional to the observed quantity, $|{\delta \u}|$.  It is
therefore convenient to define the microlensing parallax $\pi_{\rm E}
\equiv $AU$/\retilde$.

Equation (\ref{retildeeq}) relates two vectors, $\delta \u$ and
$\vect{\hat{n}} \times \vect{\hat{n}} \times \vect{a}_{\earth s}$.
But these vectors are not a priori defined on the same coordinate
axis.  The rotation between these coordinate axes, $\alpha$, is one of
the unknown quantities that must be fit in a microlensing event.
Effectively, one does not know a priori the direction of the projected
Earth-satellite separation vector in the observer $\vect{u}\retilde$
plane, though one does know its length, $|\delta \u|=|\vect{\hat{n}}
\times \vect{\hat{n}} \times \vect{a}|\pi_{\rm E}$.  Therefore, we
introduce the vector parallax $\vect{\pi}_{\rm E}$ as the quantity to
be fit.  We define $\pie$ to have length $\pi_{\rm E}$ and the same
direction as the projected Earth-satellite separation.  Thus, we have,
\begin{equation}
\label{parallax}
\delta \u = |\vect{\hat{n}} \times \vect{\hat{n}} \times
\vect{a}_{\earth s}| \vect{\pi}_{\rm E} \, .
\end{equation}

Since $\vect{a}_{\earth s}$ and $\vect{\hat{n}}$ are known, we can solve this
equation for $\vect{\pi}_{\rm E}$.  Measuring $\retilde$
is equivalent to measuring the positions of the Earth and the satellite
in the observer plane, or more precisely, their difference, $\delta \u$.

Assuming that the lens is well modeled, i.e., that $A(\u)$ is known,
the accuracy in measuring the position of the observer depends on how
rapidly the magnification varies with the position of the observer,
\begin{equation}
\frac{\sigma_{\rm phot}}{\sigma_u} \sim |\vect{\nabla} A(\u)| \, .
\end{equation}
That is, in a region where $A$ is roughly constant, one learns little
from a particular measurement, but in a region, such as the interior
approach to a caustic, where $A$ is changing rapidly, each
measurement can strongly constrain $\u$.  Note also that a single
measurement can constrain only the component of $\delta \u$ in
the direction of $\vect{\nabla} A$.  At least two measurements are
needed at different positions relative to the projected magnification pattern
to fully determine $\retilde$.

For a single lens, there is only one region where $|\vect{\nabla}{A}|$
is large, near the point caustic at the center of the lens where $A \sim
u^{-1}$ and thus $|\vect{\nabla} A| \sim u^{-2}$.  By contrast, the
binary lens has a network of caustics inside which $|\vect{\nabla} A|
\sim \Delta u^{-1.5}$ and up to 10 cusps near which $|\vect{\nabla} A|
\sim \Delta u^{-2}$.  Here, $\Delta u$ is the separation from the nearest
caustic or cusp.  

The caustics make a network of closed curves, enclosing regions in
which there are five images of the source and separating them from the
outer three-image region.  Therefore caustic crossings always occur in
pairs when the source passes into a caustic and then leaves it.  In
practice, events are not flagged as binary events until after the
first caustic crossing, so intensive monitoring by a parallax
satellite typically will not begin until then.  In general, only the
second caustic crossing will be intensively monitored, and thus can be
used to strongly constrain $\pi_{{\rm E} \perp}$, the component of
$\vect{\pi}_{\rm E}$ perpendicular to the second caustic crossing.
Completely solving the lens requires $\pi_{{\rm E} \parallel}$, the
component of parallax parallel to the second caustic crossing.

Near the second caustic crossing, the gradients of magnification are so
steep that one can even determine $\pi_{{\rm E} \perp}$ using
a terrestrial baseline of a few thousand km.  Thus, it can be measured
using two telescopes on Earth \citep{hw95,ga99}.  As these authors discuss,
three non-collinear telescopes on the Earth could completely determine
$\vect{\pi}_{\rm E}$, but in practice, it is difficult to have two
 widely separated telescopes in the southern hemisphere able to monitor
the second crossing, let alone three.

In this paper, we focus on the ability of a parallax satellite, with its
much longer baseline, to determine $\pi_{{\rm E} \parallel}$.  Such a
measurement cannot be extracted from the second caustic crossing, and
must come from other features in the magnification pattern.  If the lens is
observed near one of the other regions of high magnification, then
$\pi_{{\rm E} \parallel}$ can be measured.  For example, if the event
is caught soon enough, while the magnification is still rapidly
falling from the first caustic, the component of $\vect{\pi}_{\rm E}$
perpendicular to the first caustic crossing will be measured, which
will in general not be parallel to the second caustic crossing.  The
full microlens parallax is also determined if the source passes near a
cusp.  There is also some weak constraint from the broader part of the
magnification pattern not particularly near a caustic or cusp, where
the gradient of magnification is gentle.  The total influence of
several such regions may make a significant contribution to
$\pi_{{\rm E} \parallel}$.

We see therefore that the determination of $\retilde$, which requires
measuring both components of $\vect{\pi}_{\rm E}$, will depend in a
complicated fashion on the geometry of the lens and on the path of the
source through that geometry.  The range of events is thus best
studied by Monte Carlo simulation.

\section{Simulated Events}
\subsection{Ensemble of Microlensing Events}

We employ a Monte Carlo simulation to generate binary microlensing
events roughly as they might be in real life.  The goal here is not so
much to create an accurate model of the Galaxy and of the event
detection strategy, as to cover a variety of events.  
As the stellar density of the bulge follows a power law of $\rho
\propto r^{-1.8}$, we have for simplicity drawn sources and
lenses from a self-lensing isothermal sphere, which has the advantage
that one can analytically solve for the distribution of source and lens
distances \citep{g00b}.  The lens relative
velocity is drawn from a Maxwellian distribution with a two-dimensional
velocity dispersion of $220\, \kms$.

We choose the masses of both lenses in the binary from the ``present
day mass function'' of \citet{g00b}.  This mass function has no high
mass stars, but does include remnants such as white dwarfs and neutron
stars appropriate to an old bulge population.  Lacking further
information about the distribution of binary-lens separations, we
choose a flat distribution in $\log{d}$.  Only a narrow region of lens
separations, within a factor of two or so of $\thetae$, will
create significant caustics, so our results should be relatively
insensitive to the distribution of lens separations.

Once a lens is chosen, we randomly pick a path through the
magnification pattern,
jettisoning all events in which the path does not cross a caustic.
This path is chosen with a uniform distribution in angular impact
parameter $b \thetae$\footnote{In contrast to the usual technique for
single lenses in which events are chosen from a uniform distribution
in $b$, but are weighted towards large $\thetae$ events by multiplying
the mass function by $m^{1/2}$.}  We thus naturally weight towards lens
separations with large caustics, that is, those with large $\thetae$,
and for which the masses are separated by about $\thetae$, i.e.,
$|\vect{d}|\sim 1$.

We have assumed that observations will follow the trigger and
follow-up model that has been profitably deployed to monitor binary
lenses by several groups.  A telescope, which may or may not be one of
the two telescopes monitoring the event, finds the microlensing event
when it starts to brighten.  Once the survey telescope crosses a
caustic, the event will be classified as a binary.  This caustic
crossing will trigger follow-up by the parallax telescopes, which we
assume will begin observations 24 hours after the caustic crossing.

The time between caustic crossings (the amount of time that the
observer spends in the 5-image region, hereafter the {\it caustic
  interior time} or $\tint$) influences the ultimate
accuracy of the measurement of $\pie$: if the caustic interior time is short, there
will not be enough time to accurately measure the first caustic
crossing and use it to measure $\pi_{{\rm E}\parallel}$.  In
Figure \ref{caustimefig}, we show the range of times between caustic
crossings for both our model and those actually detected by the MACHO
and OGLE experiments towards the Galactic bulge \citep{macho00b,u00}.

\begin{figurehere}
\psfig{file=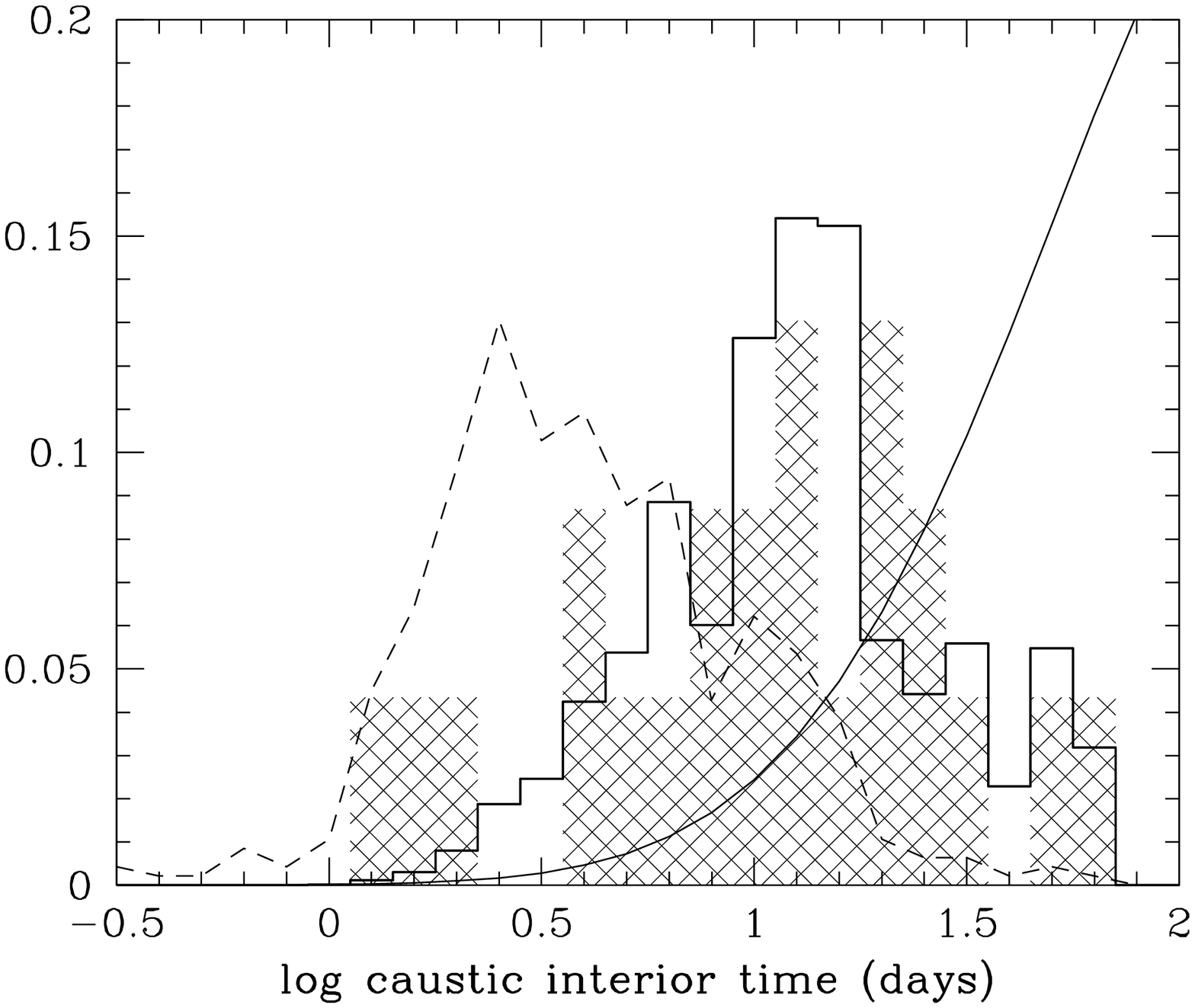, height=3.4in, width=3.4in}
\caption{
\label{caustimefig}
The dotted histogram shows caustic
interior times $\tint$ from our Monte Carlo model of binary
caustic-caustic crossing microlensing events towards the Bulge.  The
shaded histogram shows $\tint$ from the MACHO and OGLE collaborations.
The smooth curve is the OGLE single-lens detection efficiency.  The
solid histogram is our Monte Carlo model weighted by the OGLE
efficiency.  Note that it better matches the observed binary lens
events.  From this figure, we see that most caustic-crossing binary
events are missed because their caustics are too close together to
have been detected by the surveys.  }
\end{figurehere}

We see that there is a poor match between the binary events generated
by the model and those detected by the experiments: many more events
with short caustic interior times are predicted than are actually detected.
The missing ingredient is a caustic-crossing binary detection
efficiency.  If caustic interior time is short, it is possible that
this entire time will fall in a gap in the observations: there are
often gaps of days due to weather and maintenance.  Even if there are
one or two points in the caustic, it might not be unambiguously
classified as a caustic-crossing binary.

The caustic-crossing binary detection efficiency is difficult to
calculate, since it should be calculated in a manner akin to the
standard microlensing detection efficiency
\citep[e.g.,][]{machoefficiency}.  It can only be calculated after
the observations are complete and requires an extensive set of Monte Carlo
experiments on the actual data.  The efficiency will vary with the
microlensing survey program.  Such a calculation is well beyond the
scope of this work.

Detecting a binary caustic-crossing event is akin to detecting a
regular microlensing event: one needs to have a handful of
observations across the event (or across the caustic interior portion
of the binary event).  We thus expect the caustic crossing binary
detection efficiency as a function of $\tint$ to be aproximately equal
to the single lens detection efficiency for $t_{\rm E}\sim \tint$.  We
show in Figure \ref{caustimefig} how many caustic-crossing binary
events are detected in our model assuming the OGLE efficiency of
\citet{u00} can be applied to caustic-crossing binary events.

The predicted distribution is now in much better agreement with the
actually detected binary events.  The primary mismatch is the three
detected events with short $\tint$ which were all detected by the
MACHO collaboration, and are not reproduced in the efficiency-modified
model.  One of these three events is a case for which the source
passed in and out of the caustic interior twice, and would not have
been well fit without the information from the other (long) caustic
crossing.  The other two are cases for which there was rapid high
sampling frequency followup.  These three cases could not be accounted
for in the single star efficiency we adapted.  The agreement between
our efficiency-weighted model and the observed events (with the
explained exceptions for short events) suggests that our model
describes the actual caustic crossing binary events generated by the
bulge.

\subsection{Dependence on observational strategy}

Since the two telescopes monitoring the event will likely be different,
one terrestrial, and one satellite, we assume that one of the
telescopes will have a far better signal-to-noise ratio $(S/N)$ than
the other.  Our ability to measure $\vect{\pi}_{\rm E}$ will obviously scale
linearly with the $S/N$ of the weaker telescope.

If the satellite is the weaker telescope, then the photometric
observations will likely be in the source-noise dominated regime, and
the $S/N$ will then scale as $N^{-1/2}$ where $N$ is the total number
of photons collected, depending in obvious ways on the magnitude of
the source, size of the mirror, detection efficiency, pass-band, and
exposure time.  If the ground-based telescope is weaker, the source
may be in the background-noise-limited regime, depending on the size
of the seeing disk and the (magnified) brightness of the source
compared to the bulge background of 18 mag arcsec$^{-2}$.  We assume
that the observations are in the source-noise regime, and we normalize
our system to a total of 60,000 photons collected over all
exposures\footnote{To put this number into perspective, this
corresponds to the number of photons recieved from an 18th magnitude
source by the 0.9m CTIO telescope in 16 minutes of exposure.}
(ignoring magnification) or a total photometric $S/N$ of 250.

From equation (\ref{parallax}), the uncertainty in the microlens
parallax $\vect{\pi}_{\rm E}$ is inversely proportional to the projected
baseline, $|\vect{\hat{n}} \times \vect{\hat{n}} \times \vect{a}_{\earth s}|$.
We have chosen a nominal baseline $\vect{a}_{\earth s}$ of 0.2 AU at a
random orientation in the ecliptic, with the source direction
$\vect{\hat{n}}$ in Baade's Window.

We assume that the parallax observations will not begin immediately
after the first caustic crossing.  Some time will be lost waiting
for the next periodic observation, searching the data for
a caustic-crossing event (which is presently reviewed by hand before
announcing a trigger), communicating this alert to the microlensing
community, and communicating new instructions to the parallax
satellite.  We have modeled this lost time as a 24 hour delay.  This
delay is potentially serious: the first caustic crossing may be
missed.  We determine the influence of this delay by simulating continuous
observations with the delay set to 0 as well as to 24 hours.

The observational sampling rate can also be an important parameter.
We probe two different sampling regimes: continuous sampling; and
sparse sampling of once every 4 days, comparable to the typical time
between caustic crossings $\tint$ for bulge events.  Note that the
detected binary events shown in Figure \ref{caustimefig} tend to have
$\tint > 4$ due to the low efficiency for short $\tint$ events.  These
two regimes correspond respectively to what might be achieved by a
network of terrestrial telescopes combined with a dedicated satellite,
or what might be forced by a satellite that must be shared with other
programs and with scheduling determined in advance.

We have made several simplifying assumptions.  We have ignored the
parallax effect of the Earth's circular motion around the Sun,
approximating it as linear motion and we have ignored the slight
difference in velocity between the satellite and the Earth.  These
give rise to small effects that are rigorously determined by the known
motions of the Earth and satellite.  Hence, they do not affect the
error estimates relative to the naive analysis presented here.

\subsection{Analysis}

Fitting caustic-crossing binary microlensing events is still a difficult
art \citep[e.g.,][]{fitbinary}.  The sharp behavior of the caustics combined
with a highly non-linear dependence on the lens parameters yield a
complex $\chi^2$ surface.  Fortunately, we are not
concerned in this paper with finding the best fit solution to a binary lens event, but
with the precision of this solution once it is found.

In our simulation, information from the stronger telescope alone is
used to fit all the parameters of a binary microlensing event that can
be fit from a single telescope, $d,\,q,\,\rho_*,\,\thetae,\, \u_0,\,
\dot{\u}$.  Here, $q$ is the mass ratio of the binary, and $\u_0$ is
the location of the stronger telescope at some fiducial time $t_0$.
There is no published study of the ability of a single telescope to
measure these parameters for generic lenses, but experience on the few
binary lenses that have been intensively followed to date shows that
they can be fit very well.  We will assume that observations from the
stronger telescope can fit these parameters with essentially infinite
precision, and we examine the ability of the weaker telescope to fit
for the parallax.

Given a binary lens event with known parameters, we generate a time
series of photometric measurements $A_k$, each with uncertainty
$\sigma_k$.  Using the Fisher matrix technique \citep[e.g.,][]{gw96}, we
determine the covariance matrix $c_{ij}$ of the errors
\begin{equation}
c \equiv b^{-1}, \quad b_{ij} = \sum_k \sigma_k^{-2} \frac {\partial
A_k}{\partial a_i} \frac {\partial A_k}{\partial a_i} \, .
\end{equation}
Here the $a_i$ are the various parameters being fit.

We fit for the following parameters: $a_i=\{\vect{\pi}_{\rm E},F_s,F_b\}$.\
The blend flux, $F_b$, is unlensed light from a neighboring star.
This light could be from a random interloper along the line of sight,
a binary companion to the source, or from the lens itself.  Due to
blending, the measured flux is
\begin{equation}
F(\u) = F_s A(\u) + F_b \, .
\end{equation}
The source flux $F_s$ and the blend flux $F_b$ must be
independently determined at the weaker telescope except in the
unlikely case that the two telescopes have identical band passes, air
masses, and seeing conditions, which could only happen in practice if both
telescopes were satellites.

\subsection{Results}

Our results are summarized in Figure \ref{cumulative}, which shows the cumulative distributions of fractional errors,
$(\sigma_\retilde / \retilde)$ from Monte Carlo simulations of the
four cases covering a variety of sampling strategies, and testing the
effect of the caustic-crossing detection efficiency.  We show
continuous sampling beginning immediately after the first caustic
crossing, continuous sampling beginning 24 hours after the first
crossing, and sampling every four days (beginning $1-5$ days after the
first caustic crossing) including and not including the caustic
crossing detection efficiency.

\begin{figurehere}
\psfig{file=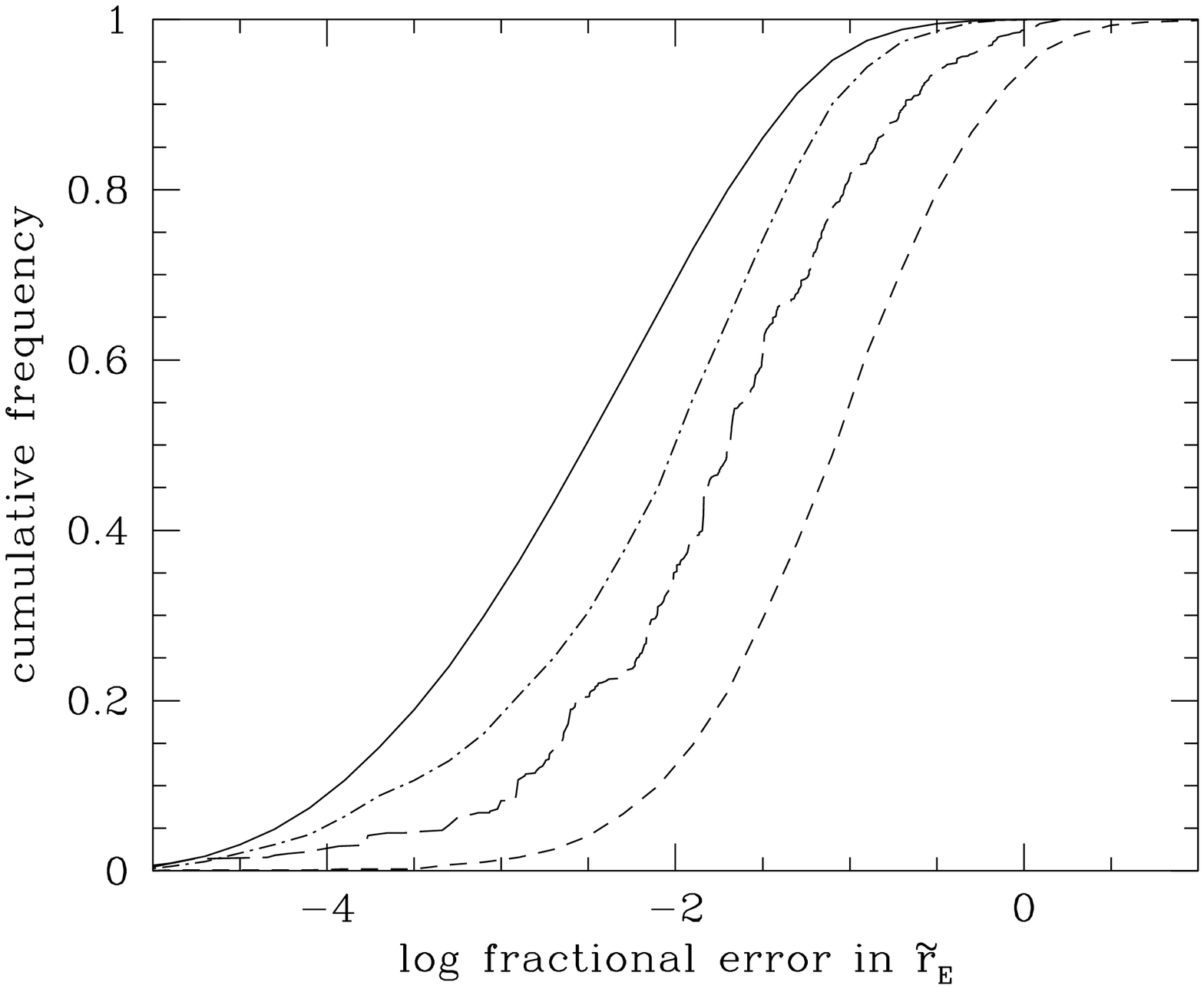, height=2.625in, width=3.4in}
\caption{
\label{cumulative}
Cumulative distributions of the fractional
error in $\retilde$ measured in a Monte Carlo sample of bulge caustic
crossing binary microlensing events.  From left to right, the curves
are for (i) no delay before beginning continuous observations, (ii) 24
hour delay before beginning continuous observations, (iii)
observations every 4 days, weighted by the caustic-crossing detection
efficiency (iv) observations every 4 days with no weighting by
efficiency.  Efficiency weighting makes no difference in the
continuous cases.  The errors are normalized assuming a total of
60,000 photons (ignoring magnification) and a baseline of 0.2 AU.
}
\end{figurehere}

These modes are listed in order of decreasing sampling agressiveness.
Note from Figure \ref{cumulative} that curves representing these
sampling strategies are arrayed from left to right: the most
aggressive sampling strategies yield the highest accuracies even
though the total telescope time is constant for these different
realizations.  The most aggressive strategy, continuous sampling
beginning immediately after the first caustic, is about 30 times more
sensitive than the least aggressive, sampling every four days
beginning $1-5$ days after the first caustic.

Sampling strategy makes such a difference because there are a few
small regions that best fix $\pie$: the areas immediately inside
caustics and immediately around cusps have the highest magnification
and strongest magnification gradients.  By sampling at a high
rate, we ensure that we catch the source while it is in these regions.
Two such regions are needed to measure both components of $\pie$.  The
strategy of beginning observations immediately after the first caustic
guarantees that both caustic crossings will be well covered.

The caustic interior time $\tint$ has a
strong effect on the fractional error in measuring the microlensing
parallax.  Events with a long $\tint$ have lower fractional error in
$\pie$ than events with short $\tint$ as shown in Figure
\ref{tintfig}.  Note from this figure that events with interior time
$\tint>10$ days all have fractional error less than $0.1$, while many
of the shorter events have fractional error larger than $0.1$.  If the
caustic interior time is short compared to the sampling frequency or
the delay from the first caustic crossing to the beginning parallax
observations, it will be difficult to determine more than one component of
$\pie$.

\begin{figurehere}
\psfig{file=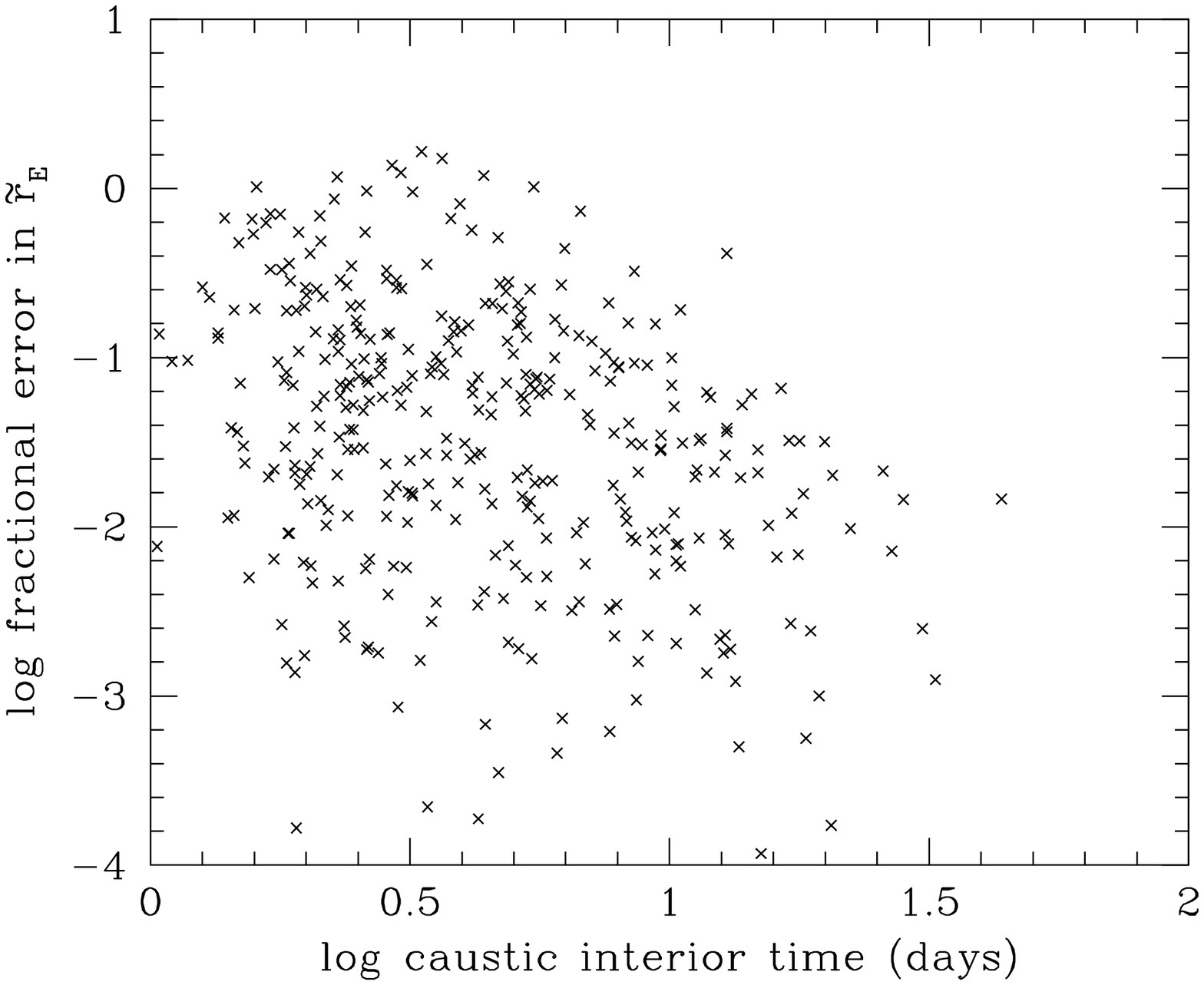, width=3.4in}
\caption{
\label{tintfig}
The fractional error in $\retilde$ plotted
against the caustic interior time $\tint$ (the time between the first
and second caustic crossings).  We have assumed sampling every four
days.  Note that events with $\tint > 10$ days have much smaller errors
than short events.  }
\end{figurehere}

We test the effect of primarily including events with long $\tint$ by
including the caustic crossing detection efficiency for one of the
curves in Figure \ref{cumulative}.  Reducing the short time events
decreases the typical fractional error by about half a dex in the case
of sampling every four days.  It has little effect on the continuous
sampling case.  There is a small inconsistency here: if microlensing
surveys are able to recognize potential binary events within 24 hours
of the first crossing, they should have a higher sensitivity to short
$\tint$ events than we have modelled for the present day surveys.
Such higher sensitivity can be achieved through aggressive groundbased
followup of potential caustic crossing events.  The MACHO
collaboration pioneered such followup, leading to the excess of short
$\tint$ events in Figure \ref{caustimefig}.

\section{Discrete degeneracies}

The Fisher matrix technique described above estimates the error in
$\retilde$ once a solution is found.  However, there may be more than
one discrete solution to a set of observations, which could foil our
ability to ultimately solve an event.  The Fisher matrix technique is
not suited to understanding the multiplicity of solutions, nor how
serious these discrete degeneracies may be.

Parallax observations of a single lens suffer from a four-fold
degeneracy.  Observations with a single telescope fix the magnitude of
that telescope's impact parameter, $|b|$.  However, each telescope could
pass on one of two sides of the lens \citep[see Fig. 2 in][]{g94}.  There
are two physically distinct interpretations to any observation: either
both telescopes pass on the same side of the lens, implying a large
$\retilde$, or they pass on opposite sides of the lens, implying a small
$\retilde$.

Fortunately, this symmetry is broken for binary lenses (except in the
special case for which the lens motion $\vect{\mu}$ is parallel to the
binary axis $\vect{d}$).  Paths on opposite sides of the center of the
lens will generate different lightcurves.  However, binary lenses can
suffer from other degeneracies.  Moderately sampled lenses can be fit
by several models \citep{d99a}, and even extremely well sampled
events can suffer from the wide-close degeneracy \citep{d99b}.
When the binary separation is wide compared to $\thetae$, the caustic
breaks up into two four-cusped caustics.  In a close binary, the
caustic breaks up into three caustics, two with three cusps and one
with four cusps.  In cases for which the binary is very wide or very
close, for a given lightcurve passing
close to or through a four-cusped caustic, there will be two
solutions, one through the four-cusped central caustic of a close
binary, and one through a four-cusped caustic from a wide binary.

For example, event MACHO 98-SMC-1 \citep{smc}, a binary
microlens in the Small Magellanic Cloud, is one of the best observed
microlensing events.  It was followed intensively by every microlensing
group, including observations every 5 minutes over the second caustic
with 1\% precision.  Despite these excellent data, there remain two
solutions, a wide solution and a close solution.

This degeneracy means only that the magnifications $A(\u)$ are similar
along the particular path of the stronger telescope through the
magnification pattern.
The weaker telescope probes a separate path through this pattern,
parallel to, but offset from the path of the stronger telescope.  The
criterion for breaking the degeneracy is that only one of the two possible
solutions, wide or close, should be able to fit the lightcurve of the
weaker telescope.

We simulate the wide-close degeneracy of event MACHO 98-SMC-1 to
determine if a parallax telescope would break the degeneracy.
Assuming that the event really was the wide solution of \citet{smc},
we generate mock data from a parallax satellite offset by
$\vect{\delta u}_w$ from the wide solution.  We then search parallax
offsets $\delta \u_c$ from the close binary solution to find the one
that best fits the lightcurve from the parallax satellite in the wide
solution.  Two such solutions are shown in Figure \ref{wideclosefig}.
We find that the difference between the two parallax satellite light
curves is small, comparable to the difference between the two
solutions from the ground.  Thus, photometry from a parallax satellite
will not break the degeneracy.

\begin{figurehere}
\psfig{file=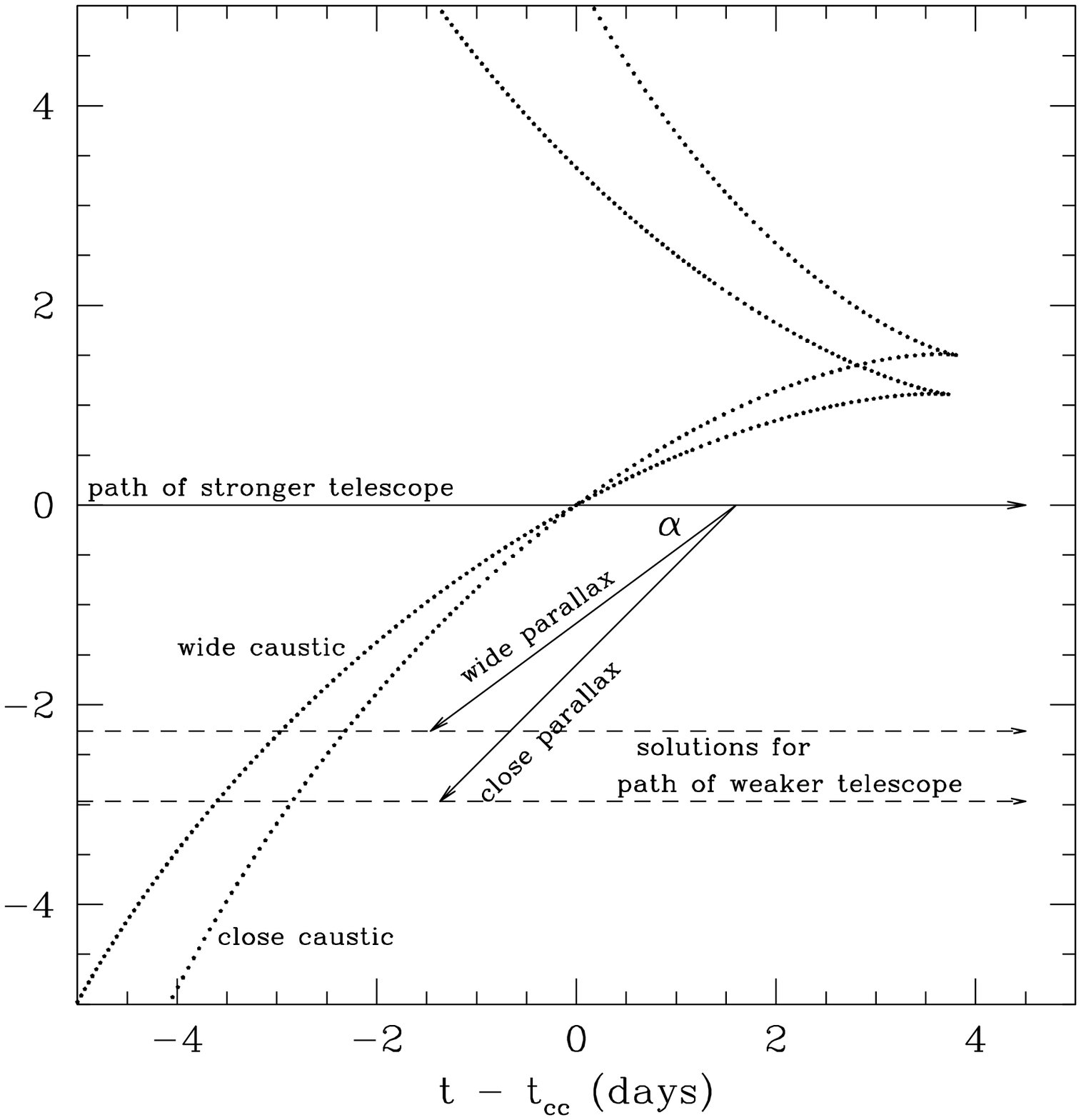, height=3.4in, width=3.4in}
\caption{
\label{wideclosefig} 
Caustics from the two solutions to degenerate event MACHO 98-SMC-1
have been rotated and scaled so that the path of the earth during
that event lies on the x axis.  The units are days from the second
caustic crossing.  Also shown are two possible degenerate solutions to
photometry from a parallax satellite.  However, the two solutions
could possibly be distinguished if the angle $\alpha$ were independently measured astrometrically.
}
\end{figurehere}

In MACHO 98-SMC-1, the caustics do not have the same shape, though in
more extreme cases such as MACHO 99-BLG-47 \citep{mb47} the caustics
almost coincide.  When the caustics from the two solutions do not
coincide, it is possible to break the degeneracy if the photometric
observations are combined with astrometric observations.  Photometric
observations of any binary microlensing event, combined with a model
of the lens, determine the orientation of the lens in solar-system
coordinates: the separations $\delta \u$ (in the lens frame) is
parallel to the separation between the two telescopes projected along
the line of sight (in the solar system frame), as described by
equation (\ref{retildeeq}).  The angle $\alpha$ between the motion of
the magnification pattern through the solar system and the projected
Earth-satellite baseline is equal to the angle between $\delta \u$ and
the source motion,\vect{\mu} (see Figure \ref{wideclosefig}).

If the caustics for the two solutions do not coincide, then $\delta
\u$ will be different for the two solutions.  For example, it is
possible that both telescopes will see the second caustic crossing at
the same time.  In that case, $\delta \u$ is parallel to the caustic.
But since the caustics in the two models do not cross the source
trajectory at the same angle, the angle between the source trajectory
and the baseline will differ between the two models.  Note that the
two parallaxes in Figure \ref{wideclosefig} are different, and most
important, have different values of $\alpha$.

However, as discussed by \citet{gh00}, the motion of the centroid of
light in the vicinity of the caustic is also degenerate: it is similar
for both the wide and close solutions.  Thus, astrometric observations
will generate the same value of $\alpha$ for both the wide and close
solutions, while as discussed above, the photmetric parallax
measurements will determine two different values of $\alpha$ depending
on the solution.  Only one of these solutions will match both the
parallax and astrometric determinations of $\alpha$.

\citet{gh00} showed that away from the caustic, the long-term
behavior of the motion of the image centroid is different for the two
solutions and can thus break the degeneracy.  The advantage of our
technique of comparing the direction of the source motion determined
both from parallax and from astrometry is that it relies only on the
astrometric observations of the caustic-crossing portion of the event,
when the event is far brighter than at baseline, allowing a great
saving in astrometric observing time.

There has, to date, been no systematic study of the wide-close
degeneracy.  We understand why it occurs in the limit of extreme-wide
and extreme close binaries \citep{d99b}.  However, we do
not know how wide or close a binary must be before it is susceptible
to this degeneracy.  Thus, we cannot tell how many of the binary events
will suffer from this degeneracy.

\section{Possible future missions}

SIM, the {\it Space Interferometry Mission}
(SIM)\footnote{http://sim.jpl.nasa.gov} is an interferometric
astrometric satellite with two effectively 25 cm mirrors, separated by
a baseline of 10 m.  One of the key projects of SIM is to follow
microlensing events; only SIM can {\it routinely} measure $\thetae$
for single lens microlens events, which it does by measuring the motion
of the centroid of the image of the source star \citep{bsvb,p98}.  SIM
will also measure $\retilde$ for these events using the same technique
proposed in this paper for measuring $\retilde$ in binary lenses
\citep{gs99}.  

SIM is at present expected to be in a trailing solar orbit, moving
away from the Earth at 0.1 AU/yr.  For present purposes, we have put
SIM at 0.2 AU.  SIM will observe bulge microlensing events in a
predetermined schedule every 4 days, and is thus our archetype of a
sparse-sampling satellite.  Microlensing events will be discovered by
a ground-based survey telescope.  Then, after a binary event crosses
the first caustic, SIM could begin monitoring this event every 4 days.

Our normalization corresponds to 1 hour of SIM observations on an
$I=18$ source.  In comparison, when studying single lenses,
\citet{gs99} assumed an $I=15$ source with 5 hours of observations, 80
times the number of photons that we have assumed here.  The typical
errors they determined for single lenses, a few percent, are
comparable to those we find here for binary lenses, but the single
lens requires two orders of magnitude more photons.

As SIM is primarily an astrometric and not a photometric
mission, it will also monitor the image centroid motion of the
microlensing event.  This motion can be used to determine $\thetae$ in
single lens events and non-caustic-crossing binary lens events, though
it is not needed for this purpose in caustic-crossing binaries since
$\thetae$ can be determined from the lightcurve alone.

Parallaxes do not have to be measured with SIM; many other satellites
could serve as well.  Consider a satellite like the GEST
satellite\footnote{http://bustard.phys.nd.edu/GEST}, but located at L2
(GEST is proposed to be in polar Earth orbit).  With its 2m telescope,
GEST would have a photometric $S/N \sim 6$ times that of SIM (for the
same exposure time) while the baseline with respect to the Earth would
be $\sim 10$ times smaller than SIM in its Earth trailing orbit.
Thus, GEST would be able to measure parallaxes with comparable
precision to SIM given the same total exposure time.  However, GEST is
proposed to continuously image $\sim 6$ fields in the bulge, so its
total exposure time would be $\sim 6$ days for a microlensing event,
144 times the 1 hour that we have assumed for SIM.  Thus, GEST could
measure parallaxes with about an order of magnitude greater accuracy
than SIM could if SIM were sampling continuously, which is another
order of magnitude better on average than SIM with sampling every 4
days.

GEST photometry would be so strong, with continuous sampling before as
well as during the event, that the ground based telescope would serve
as the weaker follow up telescope, the converse of the SIM case.
Still, given proper follow-up, much higher accuracies in measuring $\retilde$
could be achieved than in the SIM example discussed above.

\section{Discussion}

For the SIM example, our results are broadly comparable to the
accuracies derived by \citet{gs99}.  But those authors assumed a 15th
magnitude source with 5 hours of exposure time, 80 times as many
photons as we assumed here for binary lenses.  In effect, caustic
crossing binary lenses are almost two orders of magnitude more
efficient than single lenses, and three orders of magnitude more
efficient with rapid sampling.  The number of lenses studied by SIM
could be greatly increased with only a minor cost in observing time.

Since binary lenses are so much easier to study than single lenses,
many binary stellar masses could be harvested by SIM.  But are these
masses scientifically useful?  After all, SIM will measure at least
200 masses of binary stars with 1\% precision through standard
techniques.

However, standard techniques can only be applied to nearby binaries
with at least one luminous component.  Only microlensing can measure
the masses of stars in the bulge, and the masses of dark binaries
[though the masses of neutron stars and their (possibly dark)
companions can be measured in a few cases using relativistic effects
and pulsar timing \citep{tc99}].  Further, the binary masses
identified through a microlensing program will have a completely
different selection function than those observed through standard
techniques.

As we have seen, the uncertainty in a microlensing mass measurement
depends on whether or not the source crosses a caustic.  Some binary
lenses have broader caustic networks than others, with a greater
chance of crossing a pair of caustics sufficiently widely spaced to
allow an accurate parallax measurement.  For any set of lens
parameters $\{b,q,\thetae,\retilde,t_0 \}$ that will be determined
from ground-based observations, we can define the {\it efficiency of
detection} to be the fraction of paths that allow a mass measurement
of a desired accuracy.  This efficiency may be a complicated function
of the lens parameters, but it can be determined using Monte Carlo
techniques \citep[e.g.,][]{machoefficiency}.

We thank Scott Gaudi for many useful discussions.  This research was
supported by JPL contract 1226901.





\end{document}